\documentclass[aps,prb,twocolumn,groupedaddress]{revtex4}
\usepackage{graphicx}
\usepackage{bm}
\usepackage{latexsym}
\usepackage{amssymb}

\begin{document}

\title{Green's-function theory of the\\
Heisenberg ferromagnet in a magnetic field}

\author{I. Junger}
\author{D. Ihle}
\affiliation{Institut f\"{u}r Theoretische Physik,
Universit\"{a}t Leipzig, D-04109 Leipzig, Germany}
\author{J. Richter}
\affiliation{Institut f\"{u}r Theoretische Physik,
Otto-von-Guericke-Universit\"{a}t Magdeburg, D-39016 Magdeburg, Germany}
\author{A. Kl\"{u}mper}
\affiliation{Theoretische Physik, Bergische Universit\"{a}t Wuppertal,
D-42097 Wuppertal, Germany}

\date{\today}

\begin{abstract}
We present a second-order Green's-function theory of the one- and
two-dimensional $S=1/2$ ferromagnet in a magnetic field based on a decoupling of
three-spin operator products, where vertex parameters are introduced and
determined by exact relations. The transverse and longitudinal spin correlation
functions and thermodynamic properties (magnetization, isothermal magnetic
susceptibility, specific heat) are calculated self-consistently at arbitrary
temperatures and fields. In addition, exact diagonalizations on finite lattices
and, in the one-dimensional case, exact calculations by the Bethe-ansatz method
for the quantum transfer matrix are performed. A good agreement of the
Green's-function theory with the exact data, with recent quantum Monte Carlo
results, and with the spin polarization of a $\nu=1$ quantum Hall ferromagnet is
obtained. The field dependences of the position and height of the maximum in the
temperature dependence of the susceptibility are found to fit well to power
laws, which are critically analyzed in relation to the recently discussed
behavior in Landau's theory. As revealed by the spin correlation functions and
the specific heat at low fields, our theory provides an improved description of
magnetic short-range order as compared with the random phase approximation. In
one dimension and at very low fields, two maxima in the temperature dependence
of the specific heat are found. The Bethe-ansatz data for the field dependences
of the position and height of the low-temperature maximum are described by power
laws. At higher fields in one and two dimensions, the temperature of the
specific heat maximum linearly increases with the field.
\end{abstract}


\maketitle

\section{INTRODUCTION}
In the theory of low-dimensional magnetism the essential role of quantum and
thermal fluctuations, especially in the description of magnetic short-range
order (SRO) at arbitrary temperatures, is of basic interest. Whereas for
Heisenberg antiferromagnets the interplay of low dimensionality and quantum
fluctuations is important already at $T=0$, in ferromagnets quantum fluctuations
occur at non-zero temperatures only. The study of low-dimensional quantum
ferromagnets in a magnetic field was motivated by the progress in the synthesis
of new materials, such as the $\nu=1$ quantum Hall ferromagnets, \cite{MAG96}
which may be described by an effective two-dimensional (2D) $S=1/2$ Heisenberg
model, \cite{RS95,TGH98,HST00} the quasi-2D ferromagnetic insulators
A$_2$CuF$_4$ (A=K, Cs),\cite{FWS95,MKS03} La$_2$BaCuO$_5$, \cite{FWS95} and
Rb$_{2}$CrCl$_{4}$, \cite{FW98} the quasi-1D organic ferromagnet p-NPNN
(C$_{13}$H$_{16}$N$_{3}$O$_{4}$), \cite{TTN91} and the quasi-1D copper salt
TMCuC[(CH$_3$)$_4$NCuCl$_3$]. \cite{LW79}

The 2D $S=1/2$ ferromagnet in a field was investigated by Green's-function
decouplings of first order, \cite{EFJ99} i.e., by the random phase
approximation (RPA) \cite{Tja67} and the Callen decoupling, \cite{Cal63} by
Schwinger boson theories, \cite{RS95,TGH98} and by quantum Monte Carlo (QMC)
simulations. \cite{TGH98,HST00} Thereby, the magnetization
\cite{EFJ99,RS95,TGH98,HST00} and the spin lattice relaxation rate
\cite{RS95,HST00} were calculated. The 1D ferromagnet was studied by the
Bethe-ansatz method,
where some exact data for the zero-field magnetic susceptibility and specific
heat \cite{YT86} as well as for the magnetization and correlation length were
given. \cite{Tak91} Recently, in the 1D model a power law for the shift of the
temperature of the susceptibility maximum with the field was reported and
argued from Landau's theory to appear in the 2D model, too. \cite{Szn01}
Therefore, a detailed analysis of the thermodynamic quantities of the 1D and
2D Heisenberg ferromagnets as functions of temperature and field is of
interest, also for comparison with experiments.

We consider the $S=1/2$ Heisenberg model
\begin{equation}
H=-\frac{J}{2} \sum_{\langle i,j \rangle}
\bm{S}_{i} \bm{S}_{j} - h \sum_{i} S_{i}^{z}
\end{equation}
[$ \langle i,j \rangle $ denote nearest-neighbor (NN) sites; throughout we set
$J=1$] along a chain and on a square lattice. To provide an improved description
of SRO and of the thermodynamics (magnetization, magnetic susceptibility,
specific heat) as compared with the standard approaches, \cite{EFJ99} we go one
step beyond the first-order Green's-function decouplings. To this end, we adapt
the Green's-function projection method dealing with second time derivatives of
spin operators outlined in Refs.~\onlinecite{WI97, SFB00}. Furthermore, we
perform exact finite-lattice diagonalizations (ED) on an $N=16$ chain and an
$N=4 \times 4 $ square lattice using periodic boundary conditions.

The exact Bethe-ansatz results for the 1D case are obtained from an eigenvalue
analysis of the quantum transfer matrix of the Heisenberg chain, a concept
that is also the basis of the work reported in Ref.~\onlinecite{Tak91}. Here,
unlike the treatment in Ref.~\onlinecite{Tak91}, we perform this analysis by
solving a certain set of nonlinear integral equations to be found for instance
in Ref.~\onlinecite{AK98}. These integral equations were analysed in the
literature extensively for the antiferromagnetic Heisenberg chain.
The ferromagnetic case
satisfies the same set of equations with just a sign change in the temperature
dependent term. Despite this rather minor change in the analytical formulation
the numerical treatment of these equations is rather different from the
antiferromagnetic case. The iterative treatment is plagued by slow
convergence, in particular for low fields and low temperatures. A numerically
much better conditioned formulation can be derived by combining the methods of
Refs.~\onlinecite{AK98} and \onlinecite{JS99}. Details of these calculations
will be given elsewhere. Our results are in perfect agreement with those of
Refs.~\onlinecite{YT86} and \onlinecite{Tak91} if available.

\section{GREEN'S-FUNCTION THEORY}
To calculate the transverse and longitudinal spin correlation functions we
determine the two-time retarded commutator Green's functions
$
G_{\bm{q}}^{\nu \mu} (\omega)= \langle \langle S_{\bm{q}}^{\nu} ;
S_{-\bm{q}}^{\mu} \rangle \rangle_{\omega} \text{ (} \nu \mu=+-,zz\text{)}
$
 by the projection method, where we neglect the self-energy. \cite{WI97, SFB00}
Taking into account the breaking of spin-rotational symmetry by the magnetic
field we choose, as for the XXZ model, \cite{SFB00} the two-operator basis
($S_{\bm{q}}^{+},i \dot{S}_{\bm{q}}^{+}$) and ($S_{\bm{q}}^{z},i
\dot{S}_{\bm{q}}^{z}$). To approximate the time evolution of the spin operators
$ - \ddot{S}_{\bm{q}}^{+}$ and $ - \ddot{S}_{\bm{q}}^{z}$, we take the site
representation and decouple the products of three spin operators in $ -
\ddot{S}_{i}^{+}$ and $ - \ddot{S}_{i}^{z}$ along NN sequences $ \langle i,j,l
\rangle $ introducing vertex parameters $\alpha^{\nu \mu}$ in the spirit of the
scheme proposed in Refs.~\onlinecite{SFB00} and \onlinecite{ST91},
\begin{equation}
S_{i}^{+} S_{j}^{+} S_{l}^{-}= \alpha^{+-}
\langle S_{j}^{+} S_{l}^{-}\rangle S_{i}^{+} +
\alpha^{+-} \langle S_{i}^{+} S_{l}^{-}\rangle S_{j}^{+} \; ,
\end{equation}
\begin{equation}
S_{i}^{z} S_{j}^{+} S_{l}^{-}= \alpha^{zz}
\langle S_{j}^{+} S_{l}^{-}\rangle S_{i}^{z} \; .
\end{equation}
Here, following the investigation of the ferromagnet at $h=0$, \cite{ST91}
the dependence on the relative site positions of the vertex parameters
(cf. Ref.~\onlinecite{WI97}) is neglected. We obtain
\begin{equation}
- \ddot{S}_{\bm{q}}^{+}=[(\omega_{\bm{q}}^{+-})^{2}-h^{2}]
S_{\bm{q}}^{+} + 2 h i \dot{S}_{\bm{q}}^{+} \;,
\label{sekp}
\end{equation}
\begin{equation}
- \ddot{S}_{\bm{q}}^{z}=(\omega_{\bm{q}}^{zz})^{2} S_{\bm{q}}^{z} \;,
\label{sekz}
\end{equation}
with
\begin{equation}
(\omega_{\bm{q}}^{+-})^{2}=\frac{z}{2} (1-\gamma_{\bm{q}})
\{ \Delta^{+-}+2 z \alpha^{+-} C_{10} (1- \gamma_{\bm{q}}) \} \:,
\end{equation}
\begin{equation}
\Delta^{+-}=1+2 \alpha^{+-} \{(z-2) C_{11}+C_{20}-(z+1)C_{10} \} \: ;
\end{equation}
\begin{equation}
(\omega_{\bm{q}}^{zz})^{2}=\frac{z}{2} (1-\gamma_{\bm{q}})
\{ \Delta^{zz}+2 z \alpha^{zz} C_{10}^{+-} (1- \gamma_{\bm{q}}) \} \:,
\end{equation}
\begin{equation}
\Delta^{zz}=1+2 \alpha^{zz} \{(z-2) C_{11}^{+-}
+C_{20}^{+-}-(z+1)C_{10}^{+-} \} \: ,
\end{equation}
where $C_{nm}=\frac{1}{2} C_{nm}^{+-}+ C_{nm}^{zz}$, $C_{nm}^{\mu \nu}
\equiv C_{\bm{R}}^{\mu \nu} = \langle S_{0}^{\mu} S_{\bm{R}}^{\nu} \rangle $,
$\bm{R}= n \bm{e}_{x} + m \bm{e}_{y}$, $\gamma_{\bm{q}} =\frac{2}{z}
\sum_{i=1}^{z/2} \cos q_{i}$, and $z$ is the coordination number.
This approximation, Eqs.~(\ref{sekp}) and (\ref{sekz}), is equivalent to the
equation of motion decoupling in second order. \cite{ST91} Finally, we obtain
\begin{equation}
G_{\bm{q}}^{+-} (\omega)= \sum_{i=1,2}
\frac{A_{\bm{q} i}}{\omega - \omega_{\bm{q} i}}
\; ; \;\;\; \;\;\; \omega_{\bm{q} 1,2
} = h \pm \omega_{\bm{q}}^{+-} \; ,
\label{gfp}
\end{equation}
\begin{equation}
G_{\bm{q}}^{zz} (\omega)= \frac{M_{\bm{q}}^{zz}}{\omega^{2}
 - (\omega_{\bm{q}}^{zz})^{2}} \label{gfz}
\end{equation}
with
\begin{equation}
A_{\bm{q} 1,2} = \langle S^{z} \rangle \pm
\frac{1}{2 \omega_{\bm{q}}^{+-}}(M_{\bm{q}}^{+-}
- 2 h \langle S^{z} \rangle)\; .
\end{equation}
The first spectral moments $ M_{\bm{q}}^{+-}=
\langle [i \dot{S}_{\bm{q}}^{+}, S_{-\bm{q}}^{-}] \rangle $ and
$ M_{\bm{q}}^{zz}= \langle [i \dot{S}_{\bm{q}}^{z},
S_{-\bm{q}}^{z}] \rangle $ are given by the exact expressions
\begin{equation}
M_{\bm{q}}^{+-}=2 z C_{10} (1- \gamma_{\bm{q}})+ 2 h \langle S^{z} \rangle \; ,
\end{equation}
\begin{equation}
M_{\bm{q}}^{zz}=z C_{10}^{+-} (1- \gamma_{\bm{q}}) \; .
\end{equation}
The spin correlators are calculated as $C_{\bm{R}}^{\mu \nu} = \frac{1}
{N} \sum_{\bm{q}} C_{\bm{q}}^{\mu \nu} \text{e}^{i \bm{qR}}
$ with $C_{\bm{q}}^{\mu \nu} = \langle S_{\bm{q}}^{\mu}
S_{ - \bm{q}}^{\nu} \rangle $. By Eqs.~(\ref{gfp}) and (\ref{gfz}) we get
\begin{equation}
C_{\bm{q}}^{-+}=\sum_{i=1,2} A_{\bm{q} i } n(\omega_{\bm{q} i })\;, \; \; \; \;
C_{\bm{q}}^{zz}=\tilde{C}_{\bm{q}}^{zz}+ D_{\bm{q}}^{zz} \;,
\label{cqm}
\end{equation}

\begin{equation}
\tilde{C}_{\bm{q}}^{zz}=\frac{M_{\bm{q}}^{zz}}{2 \omega_{\bm{q}}^{zz}}
[1+2 n(\omega_{\bm{q}}^{zz})] \; ,
\end{equation}
where $n(\omega)=(\text{e}^{\omega/T}-1)^{-1}$. As shown in
Ref.~\onlinecite{EG79}, for the complete determination of correlation functions
calculated from commutator Green's functions one has to take into account an
additional term, if the corresponding anticommutator Green's function has a pole
at $\omega=0$. Here, we have\cite{EG79}
\begin{equation}
D_{\bm{q}}^{zz}=\lim_{\omega \rightarrow 0}
\frac{\omega}{2} G_{\bm{q}}^{(+)zz} (\omega) \; .
\label{konst}
\end{equation}
The equation of motion for the anticommutator Green's function
$G_{\bm{q}}^{(+)zz} (\omega)$ yields Eq.~(\ref{gfz}) with $M_{\bm{q}}^{zz}$ 
replaced by $M_{\bm{q}}^{(+)zz}+2 \omega C_{\bm{q}}^{zz}$, where $M_{\bm{q}}
^{(+)zz}= \langle [i \dot{S}_{\bm{q}}^{z}, S_{-\bm{q}}^{z}]_{+}
\rangle $. By the spectral theorem for $C_{\bm{q}}^{zz}$ it can be easily 
verified that $ M_{\bm{q}}^{(+)zz}=0$. Thus, Eq.~(\ref{konst}) with 
$\omega_{\bm{q}=0}^{zz}=0$ yields
\begin{equation}
D_{\bm{q}}^{zz}=C_{\bm{q}}^{zz} \delta_{\bm{q},0}=
\sum_{\bm{R}} C_{\bm{R}}^{zz} \delta_{\bm{q},0} \; .
\label{konst1}
\end{equation}
From Eqs.~(\ref{cqm}) and (\ref{konst1}) we have
$ \tilde{C}_{\bm{q}=0}^{zz}=0 $. By the relation
\begin{equation}
\frac{1}{N} \frac{\partial \langle S^{z} \rangle}{\partial h}
= \frac{1}{T} \left( \frac{1}{N} \sum_{\bm{R}} C_{\bm{R}}^{zz}-
\langle S^{z} \rangle^{2} \right) \;,
\end{equation}
following from the first and second derivatives of the partition function
with respect to $h$, in the thermodynamic limit we finally obtain
\begin{equation}
C_{\bm{R}}^{zz}=\frac{1}{N} \sum_{\bm{q} (\neq 0)} \tilde{C}_{\bm{q}}^{zz}
\text{e}^{ i \bm{q R}} + \langle S^{z} \rangle^{2} \; .
\end{equation}
Note that the transverse correlator has no additional term,
i.e.~$ D_{\bm{q}}^{-+} = 0 $, because of $ \omega_{\bm{q}=0;1,2}^{}=h \neq 0$.

The magnetization per site, $m=-2 \mu_{B} \langle S^{z} \rangle $,
is calculated as
\begin{equation}
\langle S^{z} \rangle =\frac{1}{2}-C_{0}^{-+}\;.
\end{equation}
From $ \langle S^{z} \rangle $ the isothermal magnetic susceptibility
$ \chi = 4 \mu_{B}^{2} \chi_{_S} $ with $ \chi_{_S}=
\partial \langle S^{z} \rangle / \partial h $ may be derived.

To complete our scheme, the two vertex parameters $ \alpha^{\nu \mu} (T,h)$ have
to be determined. To this end, we use the sum rule $C_{0}^{zz} =\frac{1}{4}$ and
the exact representation of the internal energy per site $u=\langle H \rangle /N 
$ in terms of $G_{\bm{q}}^{+-} (\omega)$, \cite{EG79} i.e.
\begin{eqnarray}
&-& \frac{z}{2} (C_{10}^{+-} +C_{10}^{zz}) - h \langle S^{z} \rangle = \\
&-& \frac{z}{8}- \frac{h}{2}-\frac{1}{2N} \sum_{\bm{q}}
\int_{- \infty}^{ \infty} \frac{d \omega}{2 \pi} (\varepsilon_{\bm{q}}
 + \omega) \text{Im} G_{\bm{q}}^{+-} (\omega) n(\omega) \;, \nonumber
\label{inen}
\end{eqnarray}
where $ \varepsilon_{\bm{q}} = \frac{z}{2} (1- \gamma_{\bm{q}}) + h$. Thus, we
have a closed system of equations for nine quantities ($ \langle S^{z} \rangle, 
C_{10}^{\mu \nu}, C_{11}^{\mu \nu}, C_{20}^{\mu \nu}, \alpha^{zz}, \alpha^{+-} 
$) to be determined self-consistently as functions of temperature and field. As 
may be easily seen, at $T=0$ the exact results $ \langle S^{z} \rangle 
=\frac{1}{2}$, $ C_{\bm{R}}^{zz} =\frac{1}{4}$, $ C_{\bm{R}}^{-+} =0 $ are 
reproduced.

In the case $h=0$ the spin-rotational symmetry, implying $\langle
S^{z}\rangle=0$ and $C_{\bm{R}}^{+-}=2 C_{\bm{R}}^{zz}$, is preserved by our
scheme with $\alpha^{+-}=\alpha^{zz} \equiv \alpha$, and the theory reduces to 
that of Refs.~\onlinecite{ST91} and \onlinecite{KY72}.

It is of interest to compare our results with the RPA \cite{Tja67} which
employs the decoupling $ i \dot{S}_{\bm{q}}^{+}=\omega_{\bm{q}}
S_{\bm{q}}^{+} $ and yields
\begin{equation}
G_{\bm{q}}^{+-} (\omega) = \frac{2 \langle S^{z}
\rangle}{\omega-\omega_{\bm{q}}}\;, \;\;\;
\omega_{\bm{q}}=z \langle S^{z}
\rangle (1 - \gamma_{\bm{q}}) + h \; ,
\label{rpagf}
\end{equation}
\begin{equation}
\frac{1}{ \langle S^{z} \rangle} = \frac{2}{N}
\sum_{\bm{q}} \coth \frac{\omega_{\bm{q}}}{2T} \; .
\label{rpaeq}
\end{equation}
Note that the longitudinal correlation functions cannot be obtained
by such a simple decoupling, except for $C_{10}^{zz}$ which may be
calculated in RPA by Eqs.~(\ref{inen}) and (\ref{rpagf}).

At $h=0$, in Ref.~\onlinecite{Yab91} the RPA was extended to the disordered
phase, i.e. to $T>0$ for 1D und 2D ferromagnets (Mermin-Wagner theorem).
Introducing the ratio $\displaystyle \lambda=\lim_{h \rightarrow 0} \frac{h}{z 
\langle S^{z} \rangle}$, by Eq.~(\ref{rpaeq}) with $\coth 
\omega_{\bm{q}}/2T=2T/\omega_{\bm{q}}$, $\lambda$ is calculated from
\begin{equation}
\frac{1}{N} \sum_{{\bm q}} \frac{1}{1-\gamma_{{\bm q}}+
\lambda} = \frac{z}{4 T} \; .
\label{yab}
\end{equation}
The zero-field susceptibility is given by $\chi_{_{S}} (h=0) =
(z \lambda)^{-1}$.

\section{RESULTS AND DISCUSSION}
\subsection{Magnetization}
\begin{figure}
\includegraphics{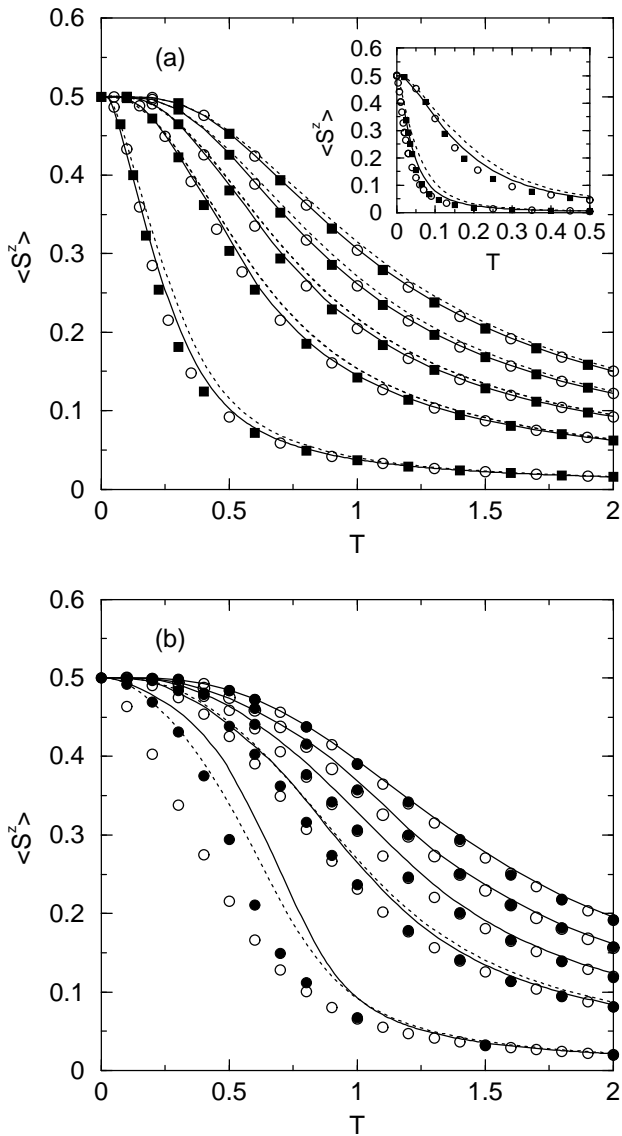}
\caption{Magnetization of the 1D (a) and 2D (b) Heisenberg ferromagnet in
magnetic fields of strengths h=1.0, 0.8, 0.6, 0.4, and 0.1, from top to bottom, 
as obtained by the Green's-function theory (solid), the ED ($\circ $), and the 
Bethe-ansatz method ($ \blacksquare $), compared with RPA results (dotted) and 
QMC data ($\bullet$, Ref.~\onlinecite{HST00}). The inset shows the low-field 
magnetization of the 1D model at $h=0.05$ and 0.005 from top to bottom.}
\label{fig_1}
\end{figure}
Considering the magnetization of the chain, in Fig.~\ref{fig_1}a the analytical
and ED results as well as our Bethe-ansatz solution are plotted and compared
with the RPA results. Let us emphasize the excellent agreement of our theory for 
the chain with the ED and Bethe-ansatz data over the whole temperature and field 
regions. For the 1D ferromagnet the RPA turns out to be a remarkably good 
approximation for $ \langle S^{z} \rangle$. In the inset the magnetization at 
low fields is depicted, since the low-field behavior of the specific heat turned 
out to be of particular interest (see below). Note that the experimental 
accessibility to the magnetic field strengths $B$ corresponding to a given $h$ 
value may be checked from the relation $h=0.116 B \text{[T]}/J \text{[meV]}$. 
Considering, e.g., the quasi-1D ferromagnet TMCuC with $J=2.6$meV, \cite{LW79} 
the value $h=0.05$ corresponds to the magnetic field $B \simeq 1$T. In 
Fig.~\ref{fig_1}b our result for the 2D ferromagnet, together with the QMC data 
for a $32 \times 32$ system, are shown. Comparing the ED with the QMC results, 
the finite-size effects are seen to be largest for low fields and at 
intermediate temperatures (cf. Ref.~\onlinecite{HST00}); for large fields ($ h 
\gtrsim 0.4$) they become small. Furthermore, as can be seen in 
Fig.~\ref{fig_1}, the finite-size effects in the 2D model are more pronounced 
than in the 1D model. This may be due to the smaller linear extension of the 2D 
system as compared with the chain of an equal number of spins. In two 
dimensions, at low fields ($ h \lesssim 0.3$) the result of our theory is 
somewhat worse than that of the RPA, although we have included additional spin 
correlations. This behavior is analogous to the results for the magnetization in 
Ref.~\onlinecite{EFJ99}, where the RPA is found to be closer to the QMC data 
than the Callen decoupling. For higher fields ($h \gtrsim 0.3 $) this tendency 
is reversed, i.e., our magnetization curves lie slightly below the RPA curves 
and closer to the QMC data.

\begin{figure}
\includegraphics{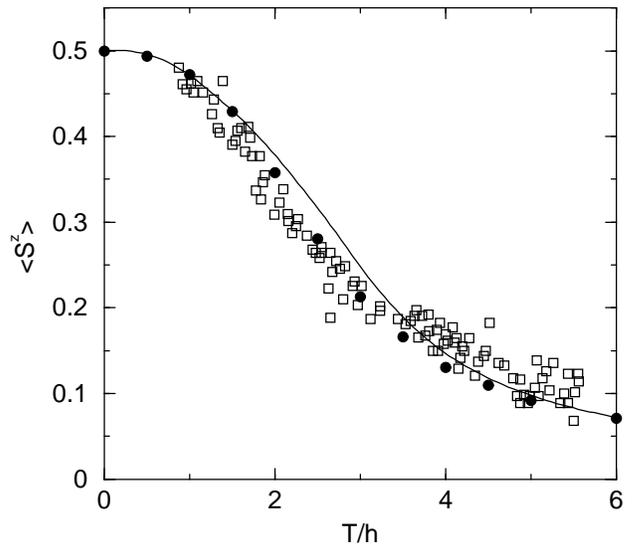}
\caption{Magnetization for the $\nu=1$ quantum Hall ferromagnet calculated at
h=0.32 (solid) in comparison with QMC ($\bullet$, Ref.~\onlinecite{HST00})
and experimental data ($\Box$, Ref.~\onlinecite{MAG96}).}
\label{fig_2}
\end{figure}
Figure \ref{fig_2} shows the spin polarization of a $\nu=1$ quantum Hall
ferromagnet measured by magnetoabsorption spectrosocopy \cite{MAG96} in
comparison with our theory and QMC data, where $h=0.32$ is
taken.\cite{TGH98,HST00} The very good agreement gives a justification for the
use of an effective 2D Heisenberg model to describe this itinerant ferromagnet.
This should be further confirmed by the comparison of other thermodynamic
quantities (magnetic susceptibility, specific heat) with experimental data which,
however, is not yet available.

\subsection{Magnetic susceptibility}
\begin{figure}
\includegraphics{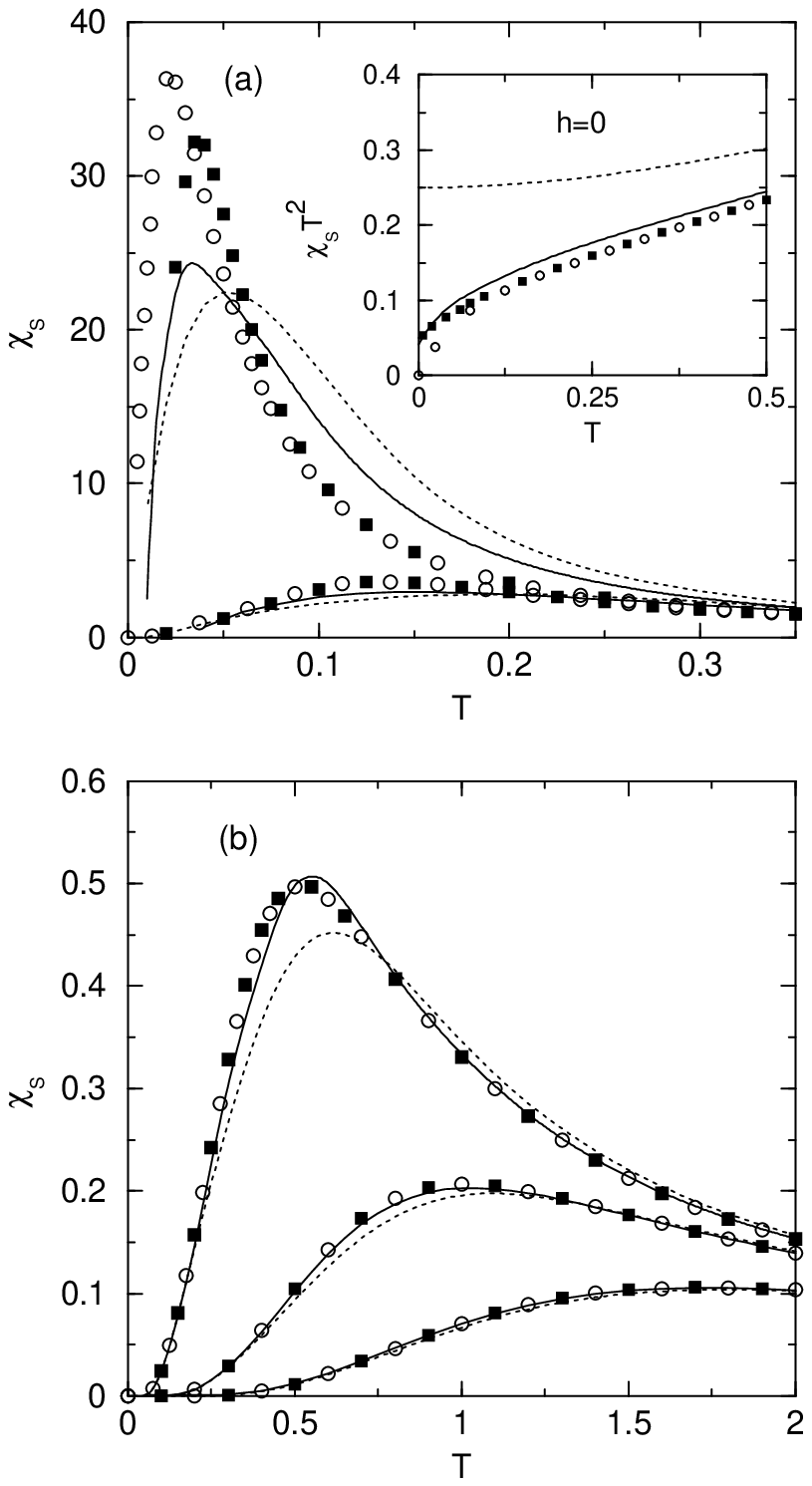}
\caption{Isothermal susceptibility of the 1D ferromagnet at low fields (a),
$h=0.005$ and 0.05, from top to bottom, and at higher fields (b), h=0.4, 1.0,
and 2.0, from top to bottom, where the Green's-function (solid), ED ($\circ$),
Bethe-ansatz ($\blacksquare$), and the RPA results (dotted) are shown. In the
inset the 1D zero-field susceptibility is depicted.}
\label{fig_3}
\end{figure}
\begin{figure}
\includegraphics{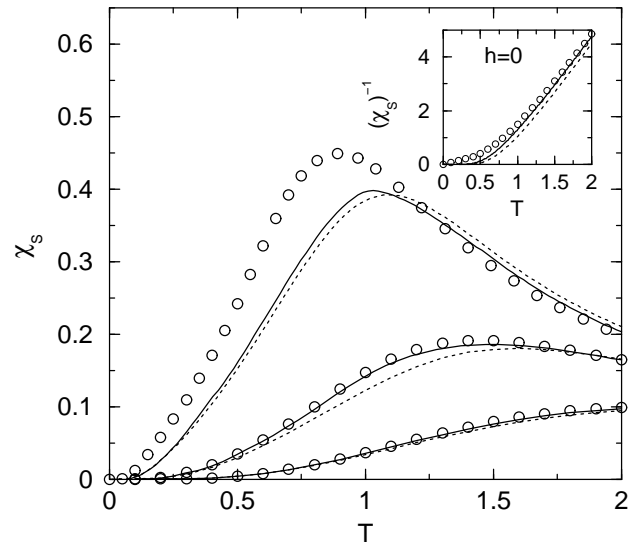}
\caption{Isothermal susceptibility of the 2D ferromagnet at h=0.4, 1.0, and 2.0,
from top to bottom. The Green's-function theory (solid) is compared with ED data
($\circ$) and RPA results (dotted). The inset shows the 2D inverse zero-field
susceptibility.}
\label{fig_4}
\end{figure}
Let us consider the isothermal susceptibility $\chi_{_{S}}=\partial \langle
S^{z} \rangle / \partial h$ shown in Figs.~\ref{fig_3} and \ref{fig_4}. For
$h=0$, $\chi_{_S}$ diverges at $T=0$ indicating the ferromagnetic phase
transition. In one dimension (see inset of Fig.~\ref{fig_3}a), the Bethe-ansatz
result $\displaystyle \lim_{T \rightarrow 0} \chi_{_{S}} T^{2}=\frac{1}{24}$
(Ref.~\onlinecite{YT86}) is reproduced by the spin-rotation-invariant
Green's-function theory. \cite{KY72} At low temperatures, the deviation of the
ED data, calculated from $\displaystyle \chi_{_{S}}= T^{-1} \sum_{\bm{R}}
C_{\bm{R}}^{zz}$, is ascribed to finite-size effects. Contrary, the RPA curve,
obtained by Eq.~(\ref{yab}) which yields $\lambda=\sqrt{1+4 T^{2}}-1$
(Ref.~\onlinecite{Yab91}), strongly deviates from the exact result. In the 2D
system, the zero-field susceptibility (see inset of Fig.~\ref{fig_4}) of the
rotation-invariant theory corrects the numerical values given in
Ref.~\onlinecite{ST91} and agrees well with the ED result. At non-zero fields we
have $\chi_{_S}(T=0)=0 $. Therefore, the susceptibility has a maximum at 
$T_{m}^{\chi} $, where $T_{m}^{\chi}$ increases and $\chi_{_S} (T_{m}^{\chi})$ 
decreases with increasing field. In one dimension (Fig.~\ref{fig_3}), the good 
agreement between Green's-function theory, Bethe-ansatz method, and ED 
corresponds to the results depicted in Fig.~\ref{fig_1}a. The deviation of our 
theory for the 2D model at $h=0.4$ and $ T \lesssim 1$ from the ED data 
(Fig.~\ref{fig_4}) is due to finite-size effects in $ \langle S^{z} \rangle $, 
as can be seen in Fig.~\ref{fig_1}b.

Recently, the field dependence of the position of the susceptibility maximum
has been discussed in connection with experiments on
La$_{0.91}$Mn$_{0.95}$O$_{3}$ showing a shift of the maximum in the temperature
derivative of the electrical resistivity at an applied field according to
$h^{2/3}$ (Ref.~\onlinecite{MRG00}). Assuming that this maximum coincides with
the maximum in the susceptibility due to spin scattering, the dependence
$T_{m}^{\chi} (h)$ was investigated in terms of Landau's theory, developed for 
anisotropic systems with $T_{c} (h=0) \neq 0 $, which yields $T_{m}^{\chi} 
\propto h^{2/3} $ (Refs.~\onlinecite{Szn01} and \onlinecite{MRG00}). Within 
Landau's theory Sznajd \cite{Szn01} claims that this power law also holds for 
isotropic ferromagnets in a field. Considering, as a further characteristics, 
the height of the susceptibility maximum $\chi_{_S} (T_{m}^{\chi})$, Landau's 
theory \cite{Szn01} yields $\chi_{_S}(T_{m}^{\chi}) \propto m^{-2} 
(T_{m}^{\chi}) \propto h^{-2/3}$. In Ref.~\onlinecite{Szn01} the isotropic spin 
chain was investigated by a real-space renormalization group method and 
$T_{m}^{\chi} \propto h^{\gamma}$ with $\gamma=0.696$ for $0.1<h<5$ was found; 
however, $ \chi_{_S}(T_{m}^{\chi})$ was not calculated.

\begin{figure}
\includegraphics{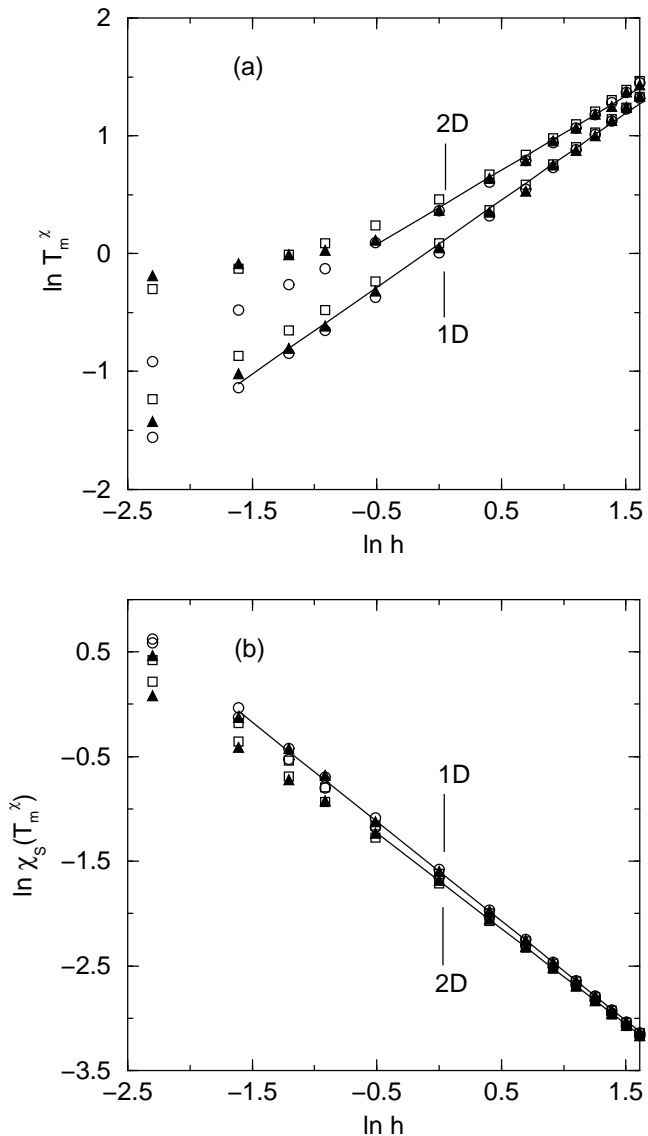}
\caption{Logarithmic plot of the field dependence of the position (a) and height 
(b) of the susceptibility maximum obtained by the Green's-function theory
($\blacktriangle$) and fit to a linear dependence (solid) in comparison with ED 
($\circ$) and RPA results ($\Box$).}
\label{fig_5}
\end{figure}
To analyze the power-law behavior in more detail, the dependence $T_{m}^{\chi}
(h)$ is plotted logarithmically in Fig.~\ref{fig_5}a for $h \geqslant 0.1$. Both
the results of our theory and the ED data for $T_{m}^{\chi}$ may be well fit to 
power laws in the 1D (2D) model for $h>0.2$ (0.6). The Green's-function theory 
yields
\begin{equation}
T_{m}^{\chi}=a \; h^{\gamma}
\label{power}
\end{equation}
with
\begin{equation}
a=\left\{ \begin{array}{l}
1.088\\
1.486
\end{array} \right.
\text{  and  }
\gamma=\left\{ \begin{array}{lll}
0.739 & ; & \text{ 1D}\\
0.620 & ; & \text{ 2D} \;.
\end{array} \right.
\label{expon}
\end{equation}
The ED results for $T_{m}^{\chi}$ (for clarity, the fit of the ED data is not
drawn) deviate only slightly from Eq.~(\ref{power}); in the 1D (2D) model we
obtain $a=1.051$ (1.460) and $\gamma=0.765$  (0.643). Considering the power-law 
behavior in the low-field region, the numerically most reliable data are 
provided by the Bethe-ansatz solution. For the 1D system at $h \leqslant 0.1$, 
the Bethe-ansatz results are described by Eq.~(\ref{power}) with
\begin{equation}
a=0.765  \text{  and  } \gamma=0.576 \;.
\label{kl_exp}
\end{equation}
Comparing our results for the 1D model with Ref.~\onlinecite{Szn01} ($\gamma
\simeq 0.7$), the $\gamma$ values are in rough agreement, whereas the absolute
values of $T_{m}^{\chi}$ found in Ref.~\onlinecite{Szn01} exceed our results by 
a factor of about 2.5.  The dependence of the maximum position on the 
dimensionality could be used for the interpretation of experimental data.

Our results for the maximum value $\chi_{_S} (T_{m}^{\chi})$ as function of $h$
in the same field region as before are plotted in Fig.~\ref{fig_5}b. Again, they
may be described by power laws for $h>0.2$ (0.6) in the 1D (2D) model. From the
Green's-function theory we obtain
\begin{equation}
\chi_{_S}(T_{m}^{\chi})=b \; h^{\beta}
\label{power2d}
\end{equation}
with
\begin{equation}
b=\left\{ \begin{array}{l}
0.202\\
0.185
\end{array} \right.
\text{  and  }
\beta=\left\{ \begin{array}{lll}
-0.951 & ; & \text{ 1D}\\
-0.914 & ; & \text{ 2D} \; .
\end{array} \right.
\label{beta_exp}
\end{equation}
The ED results for the 1D (2D) model are given by $b=0.206$ (0.191) and
$\beta=-0.964 \; \; (-0.935)$. At low fields we consider, as above, only the
Bethe-ansatz solution for the 1D model which, at $h \leqslant 0.1$, yields
Eq.~(\ref{power2d}) with
\begin{equation}
b=0.208  \text{  and  } \beta=-0.952 \;.
\label{kl_beta}
\end{equation}
Note the remarkable agreement of the Bethe-ansatz results for the field
dependence of the maximum height with the findings of the theory and
the ED data.

The results obtained for the exponent $\beta$ strongly deviate from the $\beta$
value of Landau's theory, $\beta=-\frac{2}{3}$. Moreover, we get
$m(T_{m}^{\chi}) \simeq $const. ($ \langle S^{z} \rangle (T_{m}^{\chi}) \simeq
0.3 $) which also contradicts the law $ m(T_{m}^{\chi}) \propto h^{1/3} $. This
reflects the fact that Landau's theory does not hold for 1D and 2D isotropic
ferromagnets, but is valid only on the assumption of a finite critical
temperature at $h=0$ which, however, is not realized in the 1D and 2D systems.
Therefore, the approximate agreement of the obtained $\gamma$ exponents with
$\gamma=\frac{2}{3}$ seems to be accidental.

\subsection{Specific heat}
\begin{figure}
\includegraphics{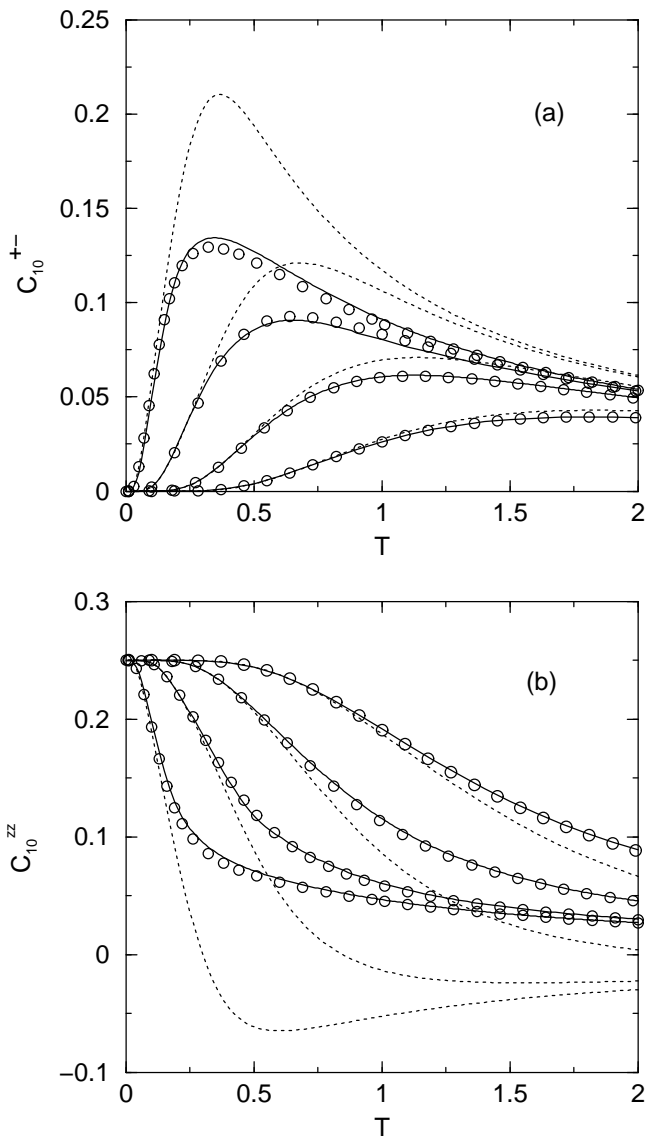}
\caption{Nearest-neighbor transverse (a) and longitudinal (b) spin correlation
functions of the 1D ferromagnet at the fields h=0.1, 0.4, 1.0, and 2.0, from
left to right. The Green's-function theory (solid) is compared with ED  
($\circ$) and RPA results (dotted).}
\label{fig_6}
\end{figure}
\begin{figure}
\includegraphics{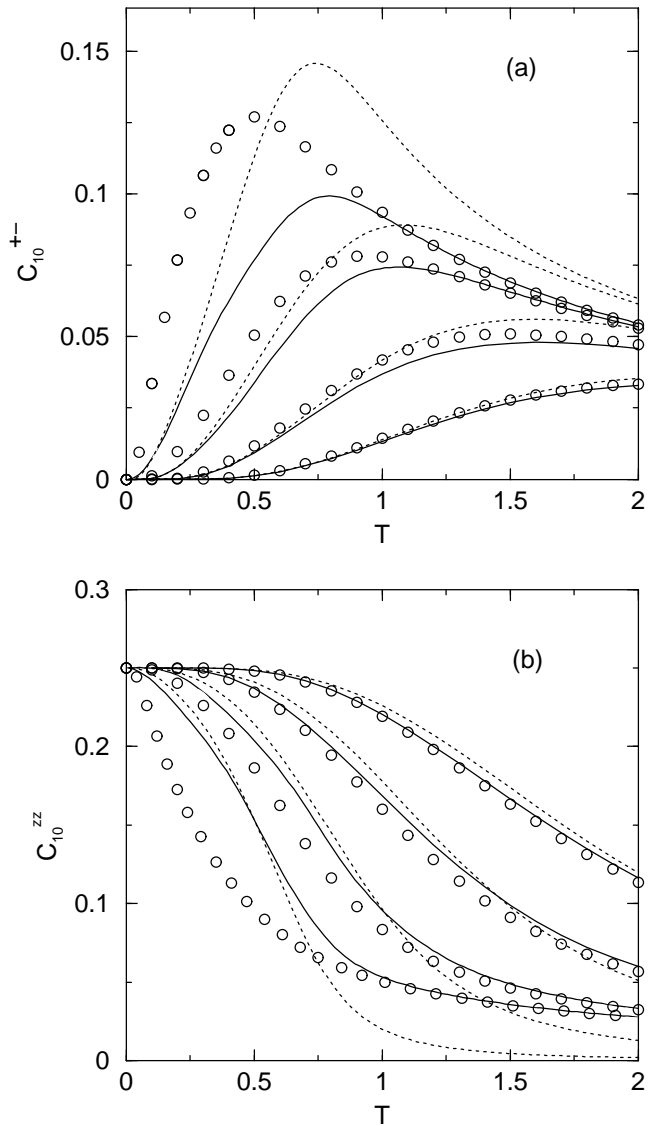}
\caption{Nearest-neighbor transverse (a) and longitudinal (b) spin correlation
functions of the 2D ferromagnet at the fields h=0.1, 0.4, 1.0, and 2.0, from
left to right. The Green's-function theory (solid) is compared with ED
($\circ$) and RPA results (dotted).}
\label{fig_7}
\end{figure}
First let us consider the NN spin correlation functions entering the internal
energy [cf. Eq.~(\ref{inen})], which are depicted for the 1D and 2D cases in
Figs.~\ref{fig_6} and ~\ref{fig_7}, respectively. In the 1D model we
obtain a very good agreement with the ED data. On the contrary, the RPA
results are unsatisfactory; in particular, the longitudinal correlators at low
fields and temperatures, obtained from the exact representation of the internal
energy, Eq.~(\ref{inen}), are negative being incompatible with theferromagnetic
SRO. In the 2D model at low fields, we ascribe the deviations of our analytical
curves from the ED data to finite-size effects; in this respect, $C_{10}^{zz}$
may be considered in analogy to $\langle S^{z} \rangle $ (Fig.~\ref{fig_1}b).

\begin{figure}
\includegraphics{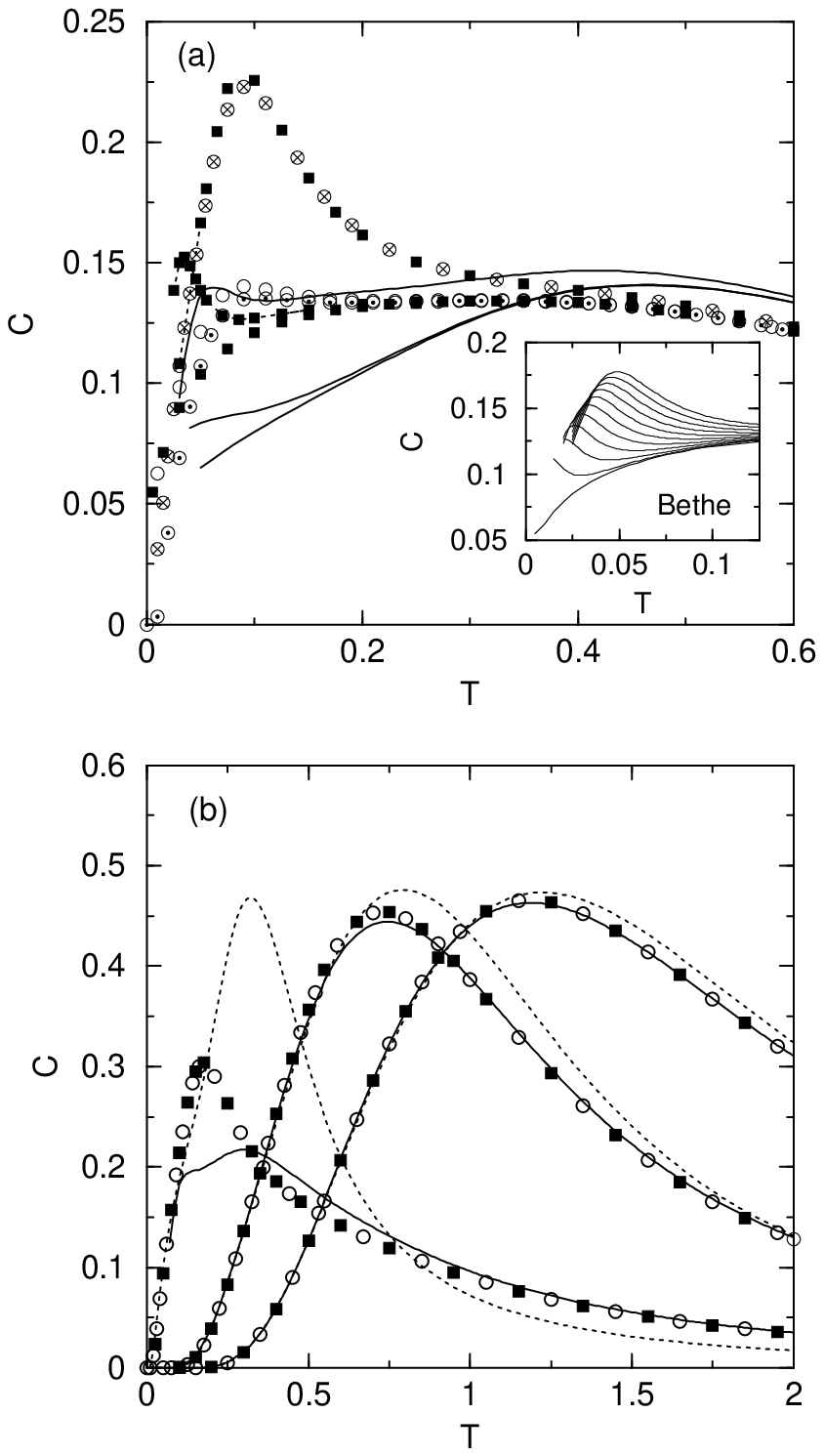}
\caption{Specific heat of the 1D ferromagnet obtained by the Green's-function
(solid) and Bethe-ansatz method ($\blacksquare$) at low fields (a), $h=0$,
0.005, and 0.03, from bottom to top, with the ED data denoted by $\odot$, 
$\circ$, and $\otimes$, respectively, and at higher fields (b), $h=0.1$, 1.0, 
and 2.0, from left to right, in comparison with ED ($\circ$) and RPA results 
(dotted). For clarity, the Bethe-ansatz data for $h=0.005$ and 0.03 at low 
temperatures are joined by dotted lines. For low fields the RPA data are not 
drawn because of the too high maximum (cf. (b)). The inset exhibits the 
Bethe-ansatz results for very low fields, $h=0$ to 0.01 in steps of 0.001, from 
bottom to top.}
\label{fig_8}
\end{figure}
Figure \ref{fig_8} displays the specific heat $C=\partial u/ \partial T$ for the
1D ferromagnet. At $h=0$, the temperature dependence of the specific heat
exhibits a broad maximum, where the value of the maximum position resulting from 
the Green's-function theory, $T_{m}^{C} (h=0) \simeq 0.45$, agrees reasonably 
well with the exact result $T_{m}^{C} (h=0) \simeq 0.35$ obtained by the 
Bethe-ansatz and ED methods. Comparing our ED data at $h=0$ with those of 
Ref.~\onlinecite{BF64} agreeing with the Bethe-ansatz results, the additional 
weak maximum at $T \simeq 0.1$ has to be ascribed to finite-size effects. The 
broad maximum and the strong decrease of the zero-field specific heat at low 
temperatures is qualitatively reproduced by the Green's-function theory, as 
already shown in Ref.~\onlinecite{KY72}. At very low magnetic fields, besides 
the high-temperature maximum, a second maximum at low temperatures develops 
which has not been reported before. The occurence of two maxima in the specific 
heat is indicated by our theory, however, at too high fields ($0.03 \lesssim h 
\lesssim 0.07$). In a detailed Bethe-ansatz analysis, two maxima in the specific 
heat are found in the field region $0 < h \lesssim 0.008$ (see inset of 
Fig.~\ref{fig_8}a). At $h \geqslant 0.008$, only one maximum appears. 
Considering the low-temperature maximum at $ h \leqslant 0.01 $, the position 
$T_{m,1}^{C}$ and height $C(T_{m,1}^{C})$ are given by the power laws
\begin{equation}
T_{m,1}^{C}=0.596 \; h^{0.542}, \; \; \; C(T_{m,1}^{C})=0.513 \; h^{0.228}.
\label{c_exp}
\end{equation}
Note that the exponent of $ T_{m,1}^{C} $ nearly agrees with that of $
T_{m}^{\chi} $ given by Eqs.~(\ref{power}) and (\ref{kl_exp}). From the
low-field specific heat it becomes evident again that our theory provides an 
improved description of SRO, as compared with RPA (cf. Fig.~\ref{fig_8}b) which 
does not yield a double maximum.

\begin{figure}
\includegraphics{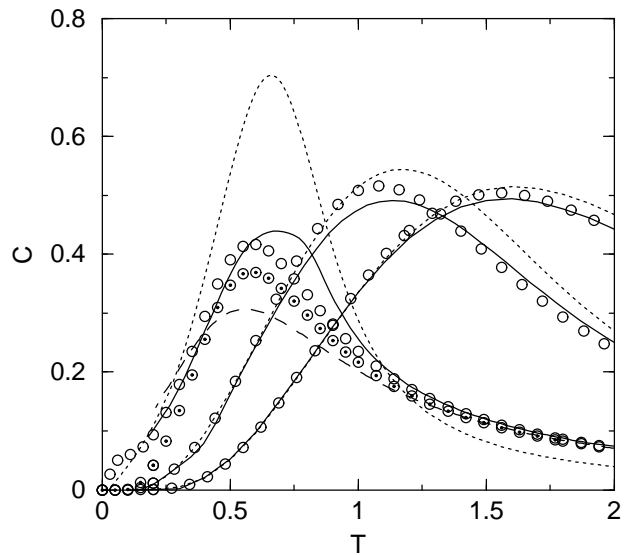}
\caption{Specific heat of the 2D ferromagnet at at $h=0.1$, 1.0, and 2.0, from
left to right, showing the Green's-function (solid), ED  ($\circ$), and the RPA
results (dotted). At $h=0$ the Green's-function theory (dashed) is compared with 
ED data ($\odot$).}
\label{fig_9}
\end{figure}
The specific heat for the 2D model is plotted in Fig.~\ref{fig_9}. We get a good
agreement with the ED results, in particular, as the position and height of the 
maximum is concerned. Note that the small low-temperature bump in the ED data 
for $h=0.1$ is a finite-size effect. The RPA curves at low fields exhibit a too 
large maximum height which we ascribe to a poor description of SRO in RPA (see 
also Fig.~\ref{fig_7}).

In the 1D and 2D systems at the fields $h>0.4$ and $h>0.1$, respectively, the
position of the specific-heat maximum obtained by the Green's-function theory 
may be described by the linear law
\begin{equation}
T_{m}^{C}=a h + b
\label{c_pow_law}
\end{equation}
with
\begin{equation}
a=\left\{ \begin{array}{l}
0.433\\
0.463
\end{array} \right.
\text{  and  }
b=\left\{ \begin{array}{lll}
0.310 & ; & \text{ 1D}\\
0.685 & ; & \text{ 2D} \;.
\end{array} \right.
\label{c_a_b}
\end{equation}

\section{SUMMARY}
In this paper we developed a Green's-function theory of the 1D and 2D $S=1/2$
Heisenberg ferromagnet in a magnetic field which goes one step further than the
RPA. The theory allows for the calculation of both transverse and longitudinal
spin correlation functions and provides an improved description of magnetic
short-range order and of the thermodynamics. Additionally, we performed exact 
finite-lattice diagonalizations on an $N=16$ chain and an $4 \times 4 $ square 
lattice and obtained exact Bethe-ansatz results for the Heisenberg chain from an 
eigenvalue analysis of the quantum transfer matrix. Stimulated by recent 
disscussions we analyzed the field dependence of the maximum in the temperature 
dependence of the isothermal magnetic susceptibility. We found power laws for 
the position and height of the susceptibility maximum which are shown not to be 
related to the predictions of Landau's theory. Paying particular attention to 
the specific heat of the Heisenberg chain, in a detailed Bethe-ansatz analysis 
the existence of two maxima in the temperature dependence  of the specific heat 
at very low magnetic fields was proven for the first time. The field dependences 
of the position and height of the low-temperature maximum obey power laws. 
Altogether, we analyzed the effects of dimensionality (1D vs 2D) on all 
thermodynamic quantities which may be relevant for the comparison with 
experiments.

\begin{acknowledgments}
The authors wish to thank N.~M.~Plakida and S.~Trimper for useful discussions.
This work was supported by the Deutsche Forschungsgemeinschaft through the
graduate college "Quantum Field Theory" (I.~J.) and the projects RI 615/12-1 and 
IH 13/7-1. We thank J. Schulenburg for assistance in ED calculations.
\end{acknowledgments}


\begin{thebibliography}{11}
\bibitem{MAG96} M.~J.~Manfra, E.~H.~Aifer, B.~B.~Goldberg, D.~A.~Broido,
L.~Pfeiffer, and K.~West, Phys.~Rev.~B \textbf{54}, R 17327 (1996).
\bibitem{RS95} N.~Read and S.~Sachdev, Phys.~Rev.~Lett.~\textbf{75},
3509 (1995).
\bibitem{TGH98} C.~Timm, S.~M.~Girvin, P.~Henelius, and A.~W.~Sandvik,
Phys.~Rev.~B \textbf{58}, 1464 (1998).
\bibitem{HST00} P.~Henelius, A.~W.~Sandvik, C.~Timm, and S.~M.~Girvin,
Phys.~Rev.~B \textbf{61}, 364 (2000).
\bibitem{FWS95} S.~Feldkemper, W.~Weber, J.~Schulenburg, and J.~Richter,
Phys.~Rev.~B \textbf{52}, 313 (1995).
\bibitem{MKS03} H.~Manaka, T.~Koide, T.~Shidara, and I.~Yamada,
Phys.~Rev.~B \textbf{68}, 184412 (2003).
\bibitem{FW98} S.~Feldkemper and W.~Weber, Phys.~Rev.~B
\textbf{57}, 7755 (1998).
\bibitem{TTN91} M.~Takahashi, P.~Turek, Y.~Nakazawa, M.~Tamura,
K.~Nozawa, D.~Shiomi, M.~Ishikawa, and M.~Kinoshita,
Phys.~Rev.~Lett.~\textbf{67}, 746 (1991).
\bibitem{LW79} C.~P.~Landee and R.~D.~Willett, Phys.~Rev.~Lett.~
\textbf{43}, 463 (1979).
\bibitem{EFJ99} A.~Ecker, P.~Fr\"{o}brich, P.~J.~Jensen, and
P.~J.~Kuntz, J.~Phys.~Condens.~Matter \textbf{11}, 1557 (1999).
\bibitem{Tja67} S.~V.~Tjablikov, in \textit{Methods in the quantum
theory of magnetism} (Plenum Press, New York, 1967).
\bibitem{Cal63} H.~B.~Callen, Phys.~Rev.~\textbf{130}, 890 (1963).
\bibitem{YT86} M.~Yamada and M.~Takahashi,
J.~Phys.~Soc.~Jpn.~\textbf{55}, 2024 (1986).
\bibitem{Tak91} M.~Takahashi, Phys.~Rev.~B \textbf{44}, 12382 (1991);
H.~Nakamura and M.~Takahashi, J.~Phys.~Soc.~Jpn.~\textbf{63}, 2563 (1994).
\bibitem{Szn01} J.~Sznajd, Phys.~Rev.~B \textbf{64}, 052401 (2001).
\bibitem{WI97} S.~Winterfeldt and D.~Ihle, Phys.~Rev.~B \textbf{56},
5535 (1997); \textbf{59}, 6010 (1999).
\bibitem{SFB00} C.~Schindelin, H.~Fehske, H.~B\"{u}ttner, and D.~Ihle,
Phys.~Rev.~B \textbf{62}, 12141 (2000);
D.~Ihle, C.~Schindelin, and H.~Fehske, Phys.~Rev.~B \textbf{64}, 054419 (2001).
\bibitem{AK98} A.~Kl\"{u}mper, Euro.~Phys.~J.~B~\textbf{5}, 677 (1998).
\bibitem{JS99} J.~Suzuki, J.~Phys.~A~\textbf{32}, 2341 (1999).
\bibitem{ST91} H.~Shimahara and S.~Takada,
J.~Phys.~Soc.~Jpn.~\textbf{60}, 2394 (1991); \textbf{61}, 989 (1992).
\bibitem{EG79} K.~Elk, W.~Gasser, in \textit{Die Methode der Greenschen
Funktionen in der Festk\"{o}rperphysik} (Akademie-Verlag, Berlin, 1979);
W.~Nolting, in \textit{Quantentheorie des Magnetismus}, vol. 2
(B.~G.~Teubner, Stuttgart, 1986).
\bibitem{KY72} J.~Kondo and K.~Yamaji, Prog.~Theor.~Phys.~\textbf{47},
807 (1972); K.~Yamaji and J.~Kondo, Phys.~Lett.~\textbf{45} A, 317 (1973).
\bibitem{Yab91} D.~A.~Yablonskiy, Phys.~Rev.~B \textbf{44}, 4467 (1991).
\bibitem{MRG00} V.~Markovich, E.~Rozenberg, G.~Gorodetsky, B.~Revzin,
J.~Pelleg, and I.~Felner, Phys.~Rev.~B \textbf{62}, 14186 (2000).
\bibitem{BF64} J.~C.~Bonner and M.~E.~Fisher,
Phys.~Rev.~\textbf{135}, A 640 (1964).
\end{thebibliography}
\end{document}